\documentclass[superscriptaddress,aps,twocolumn]{revtex4}
\usepackage{amssymb}
\usepackage{amsmath}
\usepackage{graphicx}
\usepackage{color}
\usepackage{amsfonts}
\usepackage{multirow}

\begin{document}

\title{Thermal effects on helium scattering from LiF(001) at grazing
incidence}
\author{L. Frisco}
\affiliation{Instituto de Astronom\'{\i}a y F\'{\i}sica del Espacio (UBA-CONICET).
Casilla de Correo 67, Sucursal 28, (C1428EGA) Buenos Aires, Argentina.}
\author{M.S. Gravielle\thanks{%
Author to whom correspondence should be addressed.\newline
Electronic address: msilvia@iafe.uba.ar}}
\affiliation{Instituto de Astronom\'{\i}a y F\'{\i}sica del Espacio (UBA-CONICET).
Casilla de Correo 67, Sucursal 28, (C1428EGA) Buenos Aires, Argentina.}
\date{\today }

\begin{abstract}
Grazing-incidence fast atom diffraction (GIFAD) is an exceptionally
sensitive method for surface analysis, which can be applied not only at room
temperature but also at higher temperatures. In this work we use the
He-LiF(001) system as benchmark to study the influence of temperature \ on
GIFAD patterns from insulator surfaces. Our theoretical description is based
on the Phonon-Surface Initial Value Representation (P0-SIVR) approximation, which is a semiquantum approach that includes
the phonon contribution to the elastic scattering. Within the P0-SIVR
approach the main features introduced \ by thermal lattice vibrations on the
angular distributions of scattered projectiles are investigated as a
function of the crystal temperature. We found that azimuthal and polar
spectra are strongly affected by thermal fluctuations, which modify the
relative intensities and the polar spread of the interference structures.
These findings\ are relevant for the use\ of GIFAD\ in surface research.
Moreover, the present results are contrasted with available experimental
data at room temperature.
\end{abstract}

\maketitle



\section{Introduction}

Grazing-incidence fast atom diffraction (GIFAD or FAD) is nowadays
considered as one of the most powerful nondestructive methods of surface
analysis \cite{Winter2011,Debiossac2017}. Among the attractive features of
the GIFAD technique are its extraordinary sensitivity to the morphological
and electronic characteristics of the topmost atomic layer \cite%
{Seifert2016,DelCueto2017} and the wide variety of materials that are able
to be analyzed, which ranges from insulators \cite{Schuller2012},
semiconductors \cite{Debiossac2014} and metals \cite{Rios2013} to
adsorbate-covered metal surfaces \cite{Schuller2009b}, ultrathin films \cite%
{Seifert2010}, organic-metal interfaces \cite{Seifert2013,Momeni2018}, and
graphene layers~\cite{Zugarramurdi2015}. \ In addition, even though the vast
majority of GIFAD experiments were carried out at room temperature, GIFAD
can also be applied at higher temperatures, like in the case of the
molecular beam epitaxial growth of GaAs at temperatures up to $620$ ${%
{}^{\circ }}$C, which was monitored in real time by means of GIFAD \cite%
{Atkinson2014}. Precisely, this article focuses on the influence of
temperature on GIFAD patterns, an effect that was scarcely studied in the
literature \cite{Manson2008,Aigner2008,Schuller2010,Roncin2017}.

In this paper the temperature dependence of GIFAD\ is analyzed by
considering an insulator surface - LiF - for which thermal lattice
vibrations are expected to represent the main decoherence mechanism \cite%
{Taleb2017,Schram2018}. \ In particular, we investigate thermal effects on
angular distributions of fast He atoms scattered off LiF(001) under axial
surface channeling conditions. This system has been extensively investigated
with GIFAD at room temperature \cite{Schuller2010,Schuller2007,Rousseau2007,
Schuller2008,Schuller2009,Schuller2009c,Winter2014}, becoming a prototype of
the GIFAD phenomenon. However, most of the theoretical descriptions were
based on static crystal models \cite%
{Schuller2012,Debiossac2014,Schuller2009,Gravielle2014,Muzas2017}, with the
crystal atoms at rest at their equilibrium positions, while thermal
vibration effects \ were studied in much less extent \cite%
{Manson2008,Schuller2010,Roncin2017}. Furthermore, to our knowledge there
are no available results of He-LiF GIFAD at temperatures higher than room
temperature.

To investigate thermal effects on GIFAD we make use of a recently developed
semiquantum approach, named Phonon-Surface Initial Value Representation
(P-SIVR) \cite{Frisco2019}. The P-SIVR approximation is based on the
previous SIVR approach for grazing scattering from a rigid surface \cite%
{Gravielle2014}, incorporating lattice vibrations (i.e., phonon
contributions) through a quantum description of the surface given by the
harmonic crystal model \cite{Ashcroft}. The P-SIVR probability is expanded
in terms of the number $n$ of phonons exchanged between the crystal and the
projectile during the collision. It gives rise to a series of partial P$n$%
-SIVR probabilities involving the exchange of $n$ phonons, where the
first-order term - P0-SIVR - corresponds to the elastic scattering without
net phonon exchange \cite{Frisco2019}.

P0-SIVR projectile distributions for He-LiF(001) scattering under a fixed
incidence condition are here investigated considering temperatures $T$ in
the $250-1000$ K range. With the goal of determining the contribution of
thermal lattice vibrations, P0-SIVR double differential probabilities, as a
function \ of the final azimuthal and polar angles, are contrasted with the
angular distribution for a rigid crystal, derived within the SIVR
approximation. Also azimuthal and polar spectra of scattered helium atoms
are separately analyzed as a function of $T$, finding different behaviors
along both directions. From polar P0-SIVR profiles for different $T$ values,
the lognormal dependence on the final polar angle, proposed in Ref. \cite%
{Manson2008}, is examined. Finally, the present P0-SIVR results at room
temperature are validated through the comparison with available experimental
data \cite{Schuller2009c}.

The article is organized as follows. The P0-SIVR approach is summarized in
Sec. II, while results are presented and discussed in Sec. III. In Sec. IV
we outline our conclusions. Atomic units (a.u.) are used unless otherwise
stated.

\section{Theoretical model}

Within the P0-SIVR approximation, the effective transition amplitude for
atom-surface scattering with initial (final) projectile momentum $\mathbf{K}%
_{i}$ ( $\mathbf{K}_{f}$), without net phonon exchange (i.e., with $%
K_{f}=K_{i}$ ), reads \cite{Frisco2019}
\begin{eqnarray}
\mathcal{A}^{(P0-SIVR)} &=&\int d\mathbf{R}_{o}\ f(\mathbf{R}_{o})\int d%
\mathbf{K}_{o}\ g(\mathbf{K}_{o})  \notag \\
&&\times \ \ \int d\underline{\mathbf{u}}_{o}\ a_{0}(\mathbf{R}_{o},\mathbf{K%
}_{o},\underline{\mathbf{u}}_{o}),  \label{An0}
\end{eqnarray}%
where the functions $f$ $\ $and $g$ describe the position and momentum
profiles, respectively, of the\ incident projectile wave-packet. The
function
\begin{eqnarray}
a_{0}(\mathbf{R}_{o},\mathbf{K}_{o},\underline{\mathbf{u}}_{o})
&=&\int\limits_{0}^{+\infty }dt\ \left\vert J_{P}(t)\right\vert
^{1/2}e^{i\nu _{t}\pi /2}\ \mathcal{V}_{c}(\mathbf{R}_{t})  \notag \\
&&\times \exp \left[ i\left( \varphi _{t}-\mathbf{Q}\cdot \mathbf{R}%
_{o}\right) \right]  \label{an0}
\end{eqnarray}%
represents the partial amplitude corresponding to the classical projectile
trajectory $\mathbf{R}_{t}\equiv \mathbf{R}_{t}(\mathbf{R}_{o},\mathbf{K}%
_{o},\underline{\mathbf{u}}_{o})$, which starts at the initial time $t=0$ in
the position $\mathbf{R}_{o}$ with momentum $\mathbf{K}_{o}$. This
time-dependent projectile position $\mathbf{R}_{t}$ depends on the spatial
configuration $\underline{\mathbf{u}}_{o}$ of the crystal at $t=0$, where
the underlined vector $\underline{\mathbf{u}}_{o}$ denotes the $3N$%
-dimension vector associated with the spatial displacements of the $N$ ions
contained in the crystal sample, with respect to their equilibrium
positions. In the present model such crystal deviations are considered
invariable during the collision time, which is much shorter than the
characteristic time of phonon vibrations \cite{Ashcroft}.~

In Eq. (\ref{an0}), $J_{P}(t)=\det \left[ \partial \mathbf{R}_{t}/\partial
\mathbf{K}_{o}\right] =\left\vert J_{P}(t)\right\vert \exp (i\nu _{t}\pi )$
is a Jacobian factor (a determinant), $\mathbf{Q}=\mathbf{K}_{f}-\mathbf{K}%
_{i}$ is the projectile momentum transfer, and
\begin{equation}
\varphi _{t}=\int\limits_{0}^{t}dt^{\prime }\ \left[ \frac{\left( \mathbf{K}%
_{f}-\mathbf{K}_{t^{\prime }}\right) ^{2}}{2m_{P}}-V_{PS}(\mathbf{R}%
_{t^{\prime }},\underline{\mathbf{u}}_{o})\right]  \label{fiSIVR}
\end{equation}%
is the SIVR phase at the time $t$ \cite{Gravielle2014}, where $m_{P}$ is the
projectile mass and $\mathbf{K}_{t}=m_{P}d\mathbf{R}_{t}/dt$ is the
classical projectile momentum. The potential $V_{PS}(\mathbf{R}_{t^{\prime
}},\underline{\mathbf{u}}_{o})$ represents the projectile-surface
interaction that governs the classical projectile motion, which depends on
the given spatial configuration $\underline{\mathbf{u}}_{o}$ of the crystal.
In this work, $V_{PS}$ is obtained from the pairwise additive model of Ref.
\cite{Miraglia2017}, reading

\begin{equation}
V_{PS}(\mathbf{R}_{t},\underline{\mathbf{u}})=\sum\limits_{\mathbf{r}_{%
\mathrm{B}}}v_{\mathbf{r}_{\mathrm{B}}}\left( \mathbf{R}_{t}-\mathbf{r}_{%
\mathrm{B}}-\mathbf{u}(\mathbf{r}_{\mathrm{B}})\right) ,  \label{VPS}
\end{equation}%
where $\mathbf{u}(\mathbf{r}_{\mathrm{B}})$ denotes the spatial deviation of
the crystal atom with equilibrium position $\mathbf{r}_{\mathrm{B}}$ \ and
the summation on $\mathbf{r}_{\mathrm{B}}$ covers all the occupied
Bravais-lattice sites. The potential $v_{\mathbf{r}_{_{\mathrm{B}}}}\left(
\mathbf{r}\right) $ describes the binary interaction between the projectile
and the crystal ion corresponding to the lattice site $\mathbf{r}_{_{\mathrm{%
B}}}$as a function of the relative vector $\mathbf{r}$, with $v_{\mathbf{r}_{%
\mathrm{B}}}=$ $v_{1}$ or $v_{2}$ to consider the two different ions of the
crystallographic basis. Hence, in Eq. (\ref{an0}) the crystal factor $\
\mathcal{V}_{c}(\mathbf{R}_{t})$ can be expressed as
\begin{eqnarray}
\mathcal{V}_{c}(\mathbf{R}_{t}) &=&\int d\mathbf{q}\sum\limits_{\mathbf{r}_{%
\mathrm{B}}}\tilde{v}_{\mathbf{r}_{\mathrm{B}}}(\mathbf{q})\mathbf{\exp }%
\left[ -W_{\mathbf{r}_{\mathrm{B}}}\mathbf{(q)}\right]  \notag \\
&&\times \exp \left[ i\mathbf{q}\cdot \left( \mathbf{R}_{t}-\mathbf{r}_{%
\mathrm{B}}\right) \right] ,  \label{Vcrys}
\end{eqnarray}%
with $\tilde{v}_{\mathbf{r}_{\mathrm{B}}}(\mathbf{q})$ denoting the Fourier
transform of $v_{\mathbf{r}_{_{\mathrm{B}}}}$ and $W_{\mathbf{r}_{\mathrm{B}%
}}\mathbf{(q)}$ being the momentum-dependent Debye-Waller function. This
latter function is defined as
\begin{equation}
W_{\mathbf{r}_{\mathrm{B}}}\mathbf{(q)=}\left\langle \left[ \mathbf{q}\cdot
\mathbf{u}(\mathbf{r}_{\mathrm{B}})\right] ^{2}\right\rangle /2,
\label{DWfactor}
\end{equation}%
where the dependence on $\mathbf{r}_{\mathrm{B}}$ indicates that its value
changes for the different species of the crystallographic basis, as well as
for bulk or surface positions.

The P0-SIVR probability for scattering in the direction of the solid angle $%
\Omega _{f}=(\theta _{f},\varphi _{f})$ is obtained from Eq. (\ref{An0}) as
\begin{equation}
\frac{dP^{(P0-SIVR)}}{d\Omega _{f}}=K_{f}^{2}\left\vert \mathcal{A}%
^{(P0-SIVR)}\right\vert ^{2},  \label{dP0}
\end{equation}%
where $\theta _{f}$ is the final polar angle, measured with respect to the
surface, and $\varphi _{f}$ is the azimuthal angle, measured with respect to
the axial channel. The interested reader can find the steps and assumptions
involved in the derivation of the P0-SIVR approximation in the Appendix of
Ref. \cite{Frisco2019}.

\section{Results}

In this article $^{4}$He atoms grazingly colliding with a LiF(001) surface
along the $\left\langle 110\right\rangle $ channel are used as benchmark to
investigate thermal effects on GIFAD patterns. For this purpose we applied \
the P0-SIVR approach to evaluate final helium distributions, as given by Eq.
(\ref{dP0}), considering different temperatures\ $T$ of the LiF sample.
Temperatures are confined to the $250$-$1000$ K range, for which a linear $T$%
- dependence of the mean-square vibrational amplitudes of the crystal ions
can be assumed \cite{Gupta1975}.

Along the work we kept a fixed incidence condition, given by the impact
energy $E=$ $K_{i}^{2}/(2m_{P})=1.25$ keV and the incidence angle $\theta
_{i}=1.1\deg $. (measured with respect to the surface plane). It corresponds
to the normal incidence energy $E_{\perp }=E\sin ^{2}\theta _{i}=0.46$ eV,
associated with the projectile motion perpendicular to the axial direction.
In our theoretical model we consider that before impinging on the LiF
surface, the atomic beam is collimated by a square slit of size $d$, placed
at a distance $L$ from the surface \cite{Gravielle2015}. $\ $Except for the
experimental comparison [Sec. III.D], the collimating parameters were chosen
as $d$ $=0.09$ mm and $L=36$ cm, corresponding to an extremely good
collimating condition, in accord with current experimental setups for GIFAD
\cite{Bocan2020}.

The transition amplitude $\mathcal{A}^{(P0-SIVR)}$ was calculated from Eq. (%
\ref{An0}) by using the spatial and momentum wave-packet profiles defined in
Refs. \cite{Gravielle2015,Gravielle2018}. The integral on $\mathbf{R}_{o}$
was evaluated considering that all the classical trajectories start at the
same distance from the surface, chosen as equal to the lattice constant, for
which the projectile is hardly affected by the surface interaction \cite%
{Miraglia2017}. Also the integral on \ $\mathbf{K}_{o}$ was reduced to a
two-dimensional integral over the solid angle $\Omega _{o}=(\theta
_{o},\varphi _{o})$ that determines the $\mathbf{K}_{o}$- orientation, with $%
K_{o}=K_{i}$ accounting for the negligible energy dispersion of the incident
beam \cite{Gravielle2015,Seifert2015}.

Within the P0-SIVR approach, \ thermal effects come from the integral on $%
\underline{\mathbf{u}}_{o}$, involved in Eq. (\ref{An0}), as well as from
the Debye-Waller factor $\mathbf{\exp }\left[ -W_{\mathbf{r}_{\mathrm{B}}}%
\mathbf{(q)}\right] $ which acts as an effective screening in $\mathcal{V}%
_{c}(\mathbf{R}_{t})$ [Eq. (\ref{Vcrys})]. The Debye-Waller function was
approximated as $W_{\mathbf{r}_{\mathrm{B}}}\mathbf{(q)\simeq }%
q^{2}\left\langle \mathbf{u}(\mathbf{r}_{\mathrm{B}})^{2}\right\rangle /2$,
while the integral on $\underline{\mathbf{u}}_{o}$ was evaluated with the
MonteCarlo technique by considering randomly displaced ion positions
obtained from independent Gaussian distributions with mean-square
vibrational amplitudes $\left\langle \mathbf{u}(\mathbf{r}_{\mathrm{B}%
})^{2}\right\rangle $. For the LiF crystal at a given temperature $T$, the
mean-square vibrational amplitudes $\left\langle \mathbf{u}(\mathbf{r}_{%
\mathrm{B}})^{2}\right\rangle _{T}$ were derived from the corresponding
values for the reference temperature $T_{ref}=300$ K, which were extracted
from Ref. \cite{Schuller2010}. Such reference values, which take into
account the differences between the two ionic species and between bulk and
surface (topmost layer) sites, were then extrapolated as a function of $T$
following the temperature dependence given in Ref. \cite{Gupta1975}. That
is, we approximate $\left\langle \mathbf{u}(\mathbf{r}_{\mathrm{B}%
})^{2}\right\rangle _{T}$ $\approx \lbrack 1+B(\mathbf{r}_{\mathrm{B}%
})(T/T_{ref}-1)]\left\langle \mathbf{u}(\mathbf{r}_{\mathrm{B}%
})^{2}\right\rangle _{T_{ref}}$, with $B(\mathbf{r}_{\mathrm{B}})=0.795$ ($%
0.890$) for Li (F) ions.

\subsection{Thermal effects on the $(\protect\theta _{f},\protect\varphi %
_{f})$ distributions}

\begin{figure}[tbp]
\includegraphics[width=0.5 \textwidth]{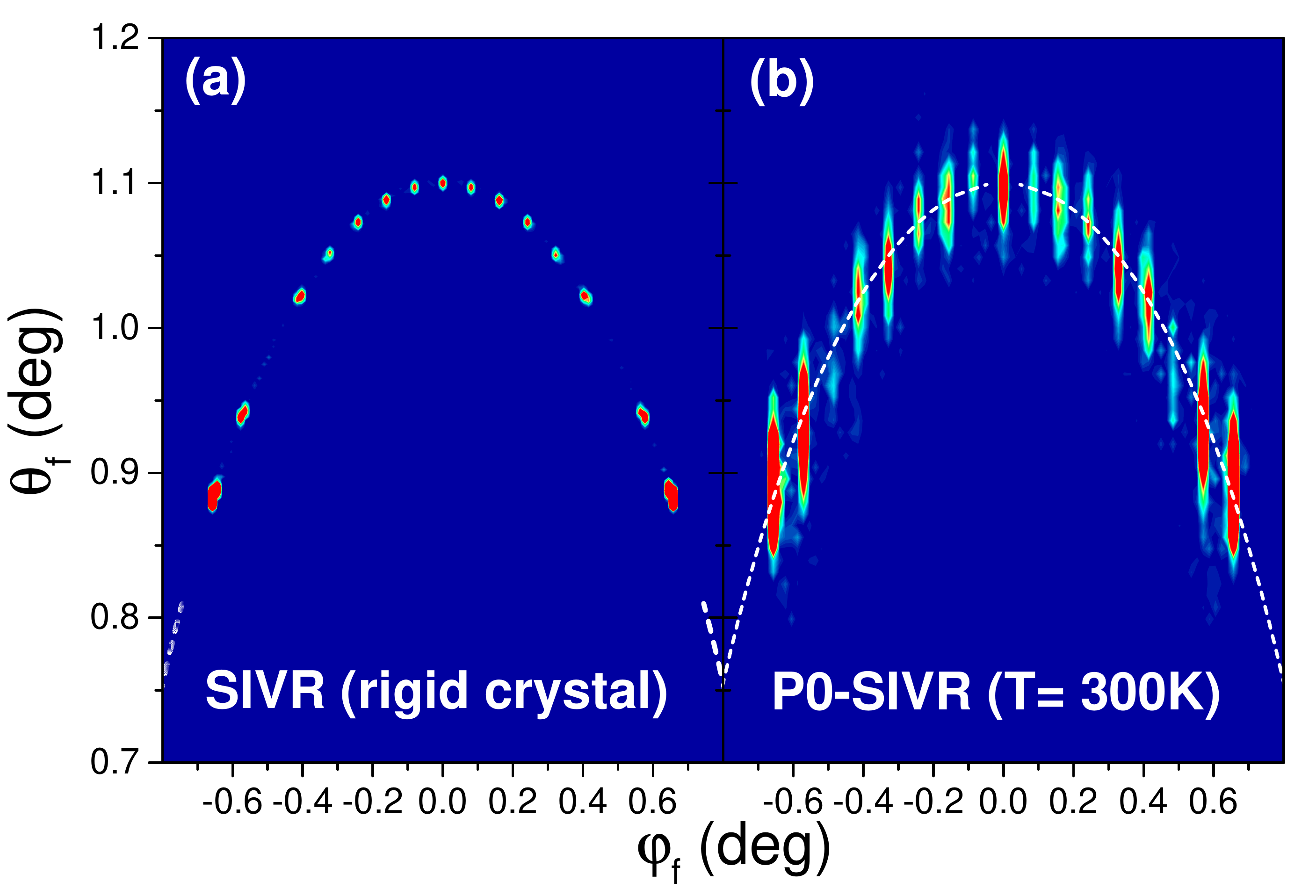} \centering
\caption{(Color online) Two-dimensional projectile distributions, as a
function of $\protect\theta _{f}$ and $\protect\varphi _{f}$, for $1.25$ keV
$^{4}$He atoms scattered off LiF(001) at the temperature $T=300$ K.
Incidence along the $\left\langle 110\right\rangle $ channel with the
grazing angle $\protect\theta _{i}=1.1\deg $ is considered. Results derived
within (a) the SIVR approximation, for a rigid crystal, and (b) the P0-SIVR
approach, including thermal vibrations, are displayed.}
\label{map-0t300}
\end{figure}

Since GIFAD experiments involving insulator surfaces are usually carried out
at room temperature, we start analyzing thermal effects at $T=300$ K. In
Fig. \ref{map-0t300} the P0-SIVR two-dimensional (2D) distribution as a
function of the final scattering angles $\theta _{f}$ and $\varphi _{f}$,
for a lithium fluoride surface at $T=300$ K, is displayed along with the
angular distribution derived within the SIVR approach, \ which assumes an
ideal LiF crystal with its ions at rest at their equilibrium positions \cite%
{Gravielle2014}.\ While the SIVR\ distribution [Fig. \ref{map-0t300} (a)]
presents nearly circular spots associated with equally $\varphi _{f}$-
spaced Bragg maxima (with order $m$, $m=0,\pm 1,\pm 2,..$), the phonon
contribution included in the P0-SIVR approximation transforms such Bragg
peaks into elongated strips, as shown in Fig. \ref{map-0t300} (b). Moreover,
in absence of lattice vibrations the Bragg maxima of Fig. \ref{map-0t300}
(a) lie on a circle of radius $\theta _{i}$ (the Laue circle), which is a
sign of elastic scattering from an ideal surface under extremely good
collimating conditions \cite{Gravielle2015,Debiossac2016}. But when thermal
fluctuations of the crystal lattice are taken into account, as it happens in
the P0-SIVR distribution of Fig. \ref{map-0t300} (b), the maximum intensity
of some Bragg orders appears at a polar angle slightly shifted above or
below the Laue circle, as it is usually observed in GIFAD experiments \cite%
{Winter2011}.

\begin{figure}[tbh]
\includegraphics[width=0.4\textwidth]{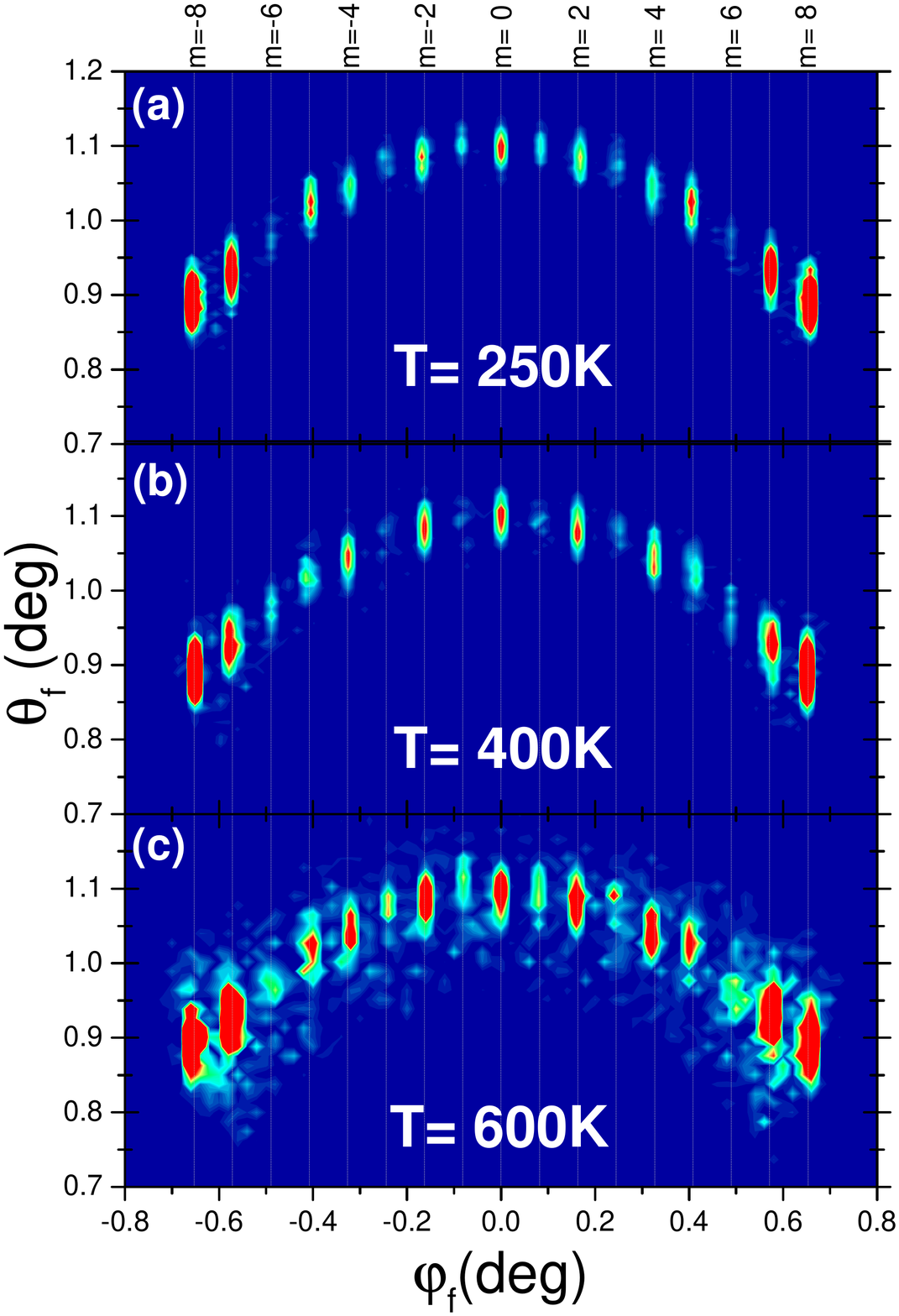} \centering
\caption{(Color online) Analogous to Fig. \protect\ref{map-0t300} (b) for
three different temperatures: (a) $T=250$ K, (b) $T=400$ K, and (c) $T=600$
K. Vertical dashed lines, ideal Bragg-peak positions with their orders $m$
indicated in the upper axis.}
\label{map-t250vst400vst600}
\end{figure}

In order to investigate how the previous effects change with the
temperature, in Fig. \ref{map-t250vst400vst600} we display 2D angular
distributions derived with the P0-SIVR approach for LiF crystals at
different temperatures: (a) $T=250$ K, (b) $T=400$ K, and (c) $T=600$ K. In
every panel, the intensity scale was normalized at the maximum intensity of
the GIFAD distribution, which corresponds to the outermost peak associated
with rainbow scattering. From Fig. \ref{map-t250vst400vst600} it is evident
that the azimuthal positions of the Bragg maxima are independent of $T$,
being completely determined by the crystallographic parameters of the ideal
surface \cite{Winter2011}. Furthermore, the three P0-SIVR distributions of
Fig. \ref{map-t250vst400vst600} look similar to each other, displaying
intense $m=0$ \ (central) and $m=\pm 2$ maxima, along with almost suppressed
$m=\pm 1$, $\pm 3$ and $\pm 6$ peaks. But in spite of this overall
similitude, we found that the relative intensities as well as the polar
spread of the Bragg maxima depend on the temperature, this latter increasing
as $T$ augments. Note that GIFAD structures are clearly visible even for a
temperature as high as $T=600$ K, although they start to blur at this
temperature. This fact is in accord with the experiments for semiconductor
surfaces \cite{Debiossac2014,Atkinson2014}, where GIFAD patterns were
observed at high temperatures.

\begin{figure}[tbh]
\includegraphics[width=0.4\textwidth]{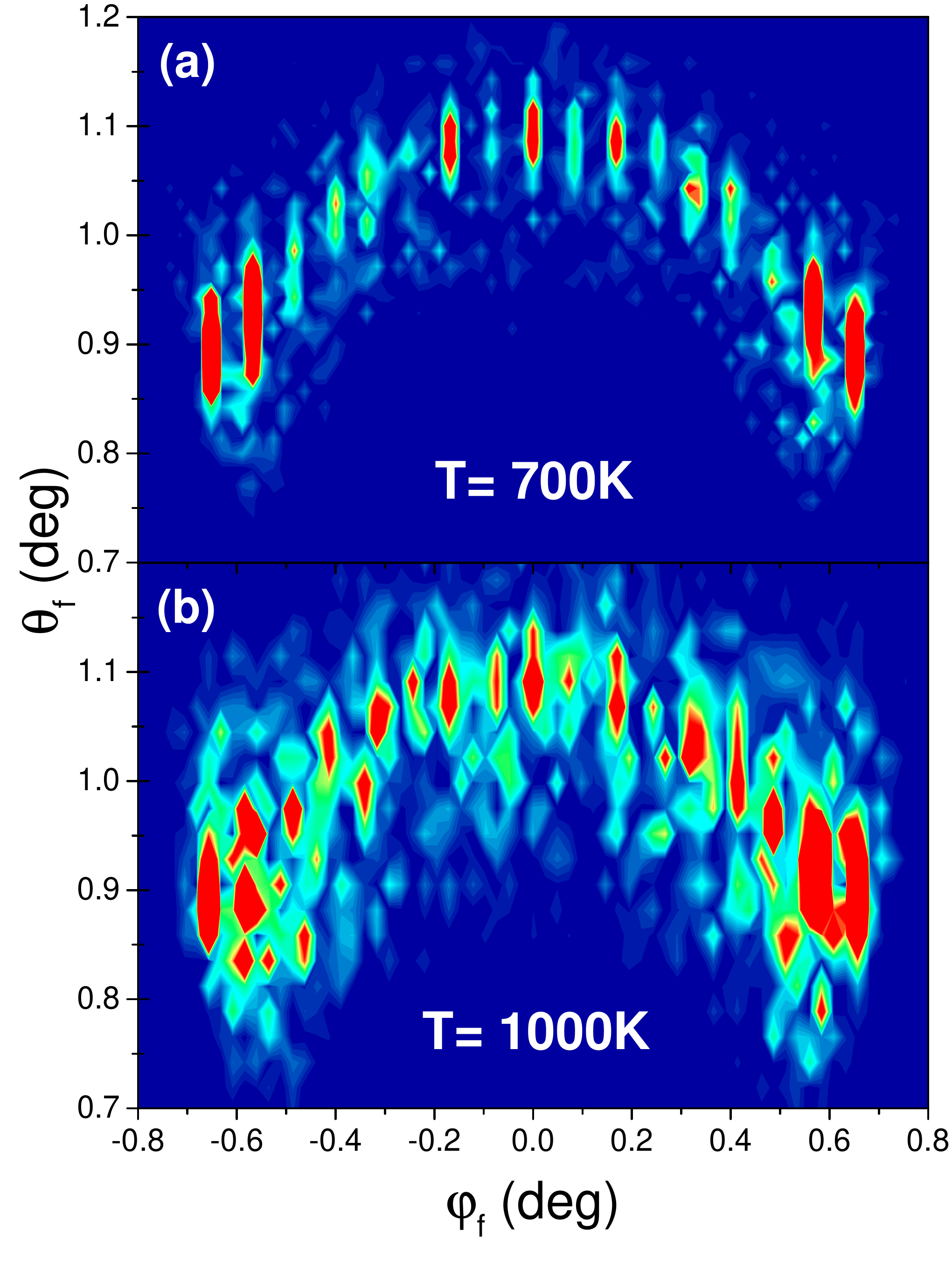} \centering
\caption{(Color online) Analogous to Fig. \protect\ref{map-0t300} (b) for
(a) $T=700$ K and (b) $T=1000$ K.}
\label{map-t700vst1000}
\end{figure}

For the He-LiF(001) system, GIFAD patterns gradually smudge as the
temperature rises above $700$ K, ending up almost completely blurred for LiF
at $T=1000$ K, for which thermal vibrations strongly deteriorate the
coherence, as observed in Fig. \ref{map-t700vst1000}. Besides, the intensity
of the interference structures decreases sharply as the temperature
increases, which might contribute to making their experimental detection
difficult in this $T$- range.

A more in-depth inspection of the aforementioned thermal effects along the
azimuthal and polar directions is presented in Secs. III.B and III.C,
respectively.

\subsection{Influence of $T$ on the azimuthal spectra}

\begin{figure}[tbp]
\includegraphics[width=0.5\textwidth]{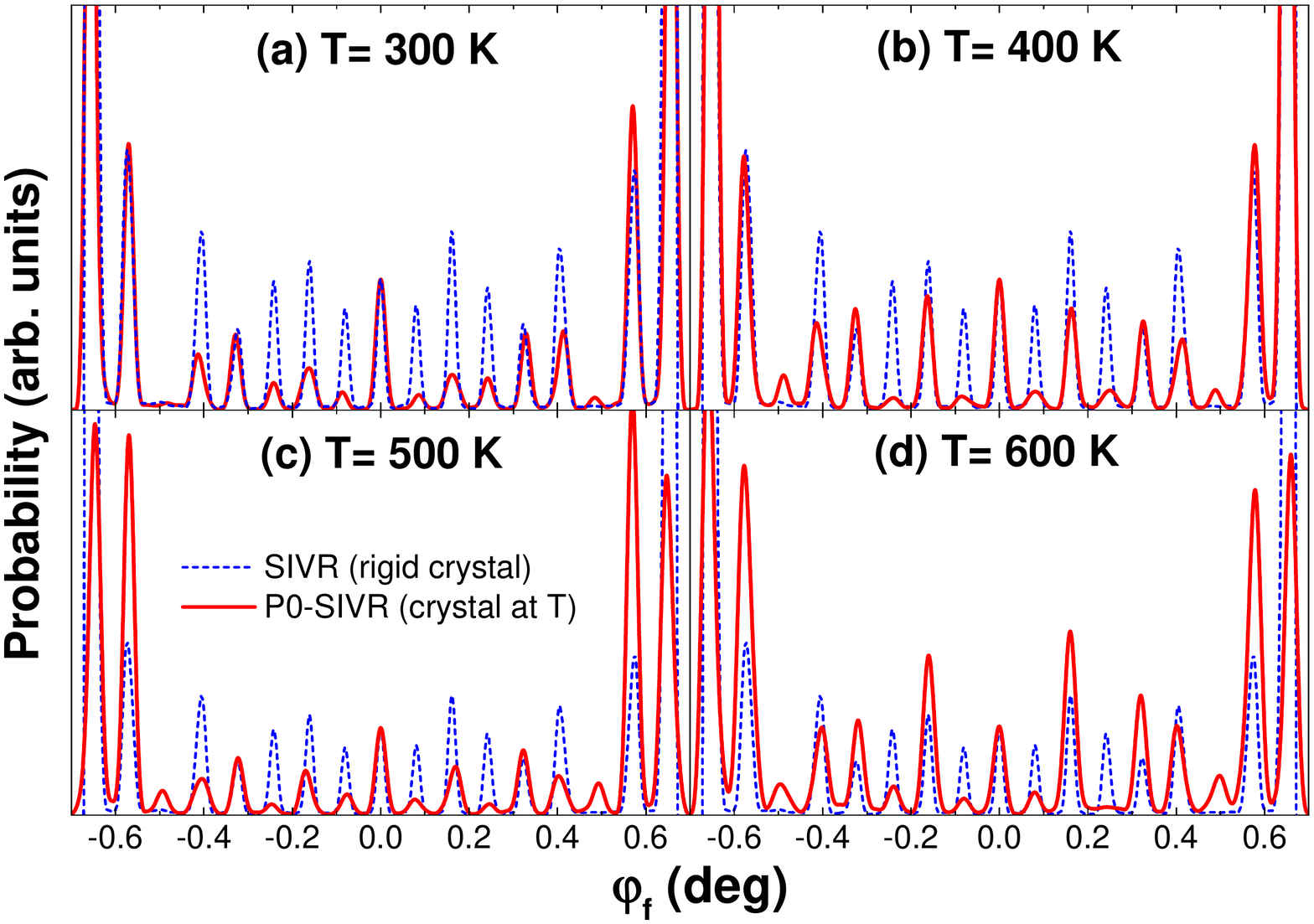} \centering
\caption{(Color online) Differential probability $dP^{(P0-SIVR)}/d\protect%
\varphi _{f}$, as a function of the azimuthal angle $\protect\varphi _{f}$,
for the case of Fig. \protect\ref{map-0t300} considering different crystal
temperatures: (a) $T=300$ K, (b) $T=400$ K, (c) $T=500$ K, (d) $T=600$ K. In
all the panels, red solid line, P0-SIVR probability including thermal
vibrations; blue dashed line, SIVR probability for a rigid crystal.}
\label{spectra-T}
\end{figure}

Surface characterization by means of GIFAD is commonly based on the
theory-experiment comparison of the relative intensities of the interference
maxima along the transverse direction, perpendicular to the incidence
channel \cite{Schuller2012,Debiossac2014}. Therefore, to use GIFAD as a
surface analysis tool it is important to know the influence of temperature
on such transverse spectra, i.e., on the azimuthal projectile distributions.
In Fig. \ref{spectra-T} we plot $dP^{(P0-SIVR)}/d\varphi _{f}$, as a
function of the azimuthal angle $\varphi _{f}$, for temperatures varying
between $300$ and $600$ K. These single differential probabilities were
calculated by integrating Eq. (\ref{dP0}) over a reduced annulus of mean
radius $\theta _{i}$ and central thickness $0.03\deg $, as it is usually
done to derive the experimental projected intensities \cite%
{Winter2011,Debiossac2016}. In all the panels P0-SIVR results including
thermal lattice fluctuations are contrasted with the azimuthal distribution
for an ideal rigid LiF crystal, derived within the SIVR approach,
normalizing both spectra at $\varphi _{f}=0$. From Fig. \ref{spectra-T} we
confirm that not only the $\varphi _{f}$- positions of the Bragg peaks are
independent of $T$, but also the azimuthal widths of these maxima are weakly
affected by thermal vibrations, being mainly determined by the number of
parallel channels coherently illuminated by the atomic beam \cite%
{Seifert2015,Gravielle2018}. On the contrary, the relative intensities of
the Bragg peaks are strongly affected by the lattice fluctuations, which
strongly suppress the intensities corresponding to the $m=\pm 1$ and $\pm 3$
Bragg orders, this fact being observed for all the $T$ values. Therefore,
this suppression effect might be used to determine the lattice vibration
contribution in the studied case.

\subsection{Influence of $T$ on the polar profiles}

\begin{figure}[tbp]
\includegraphics[width=0.5\textwidth]{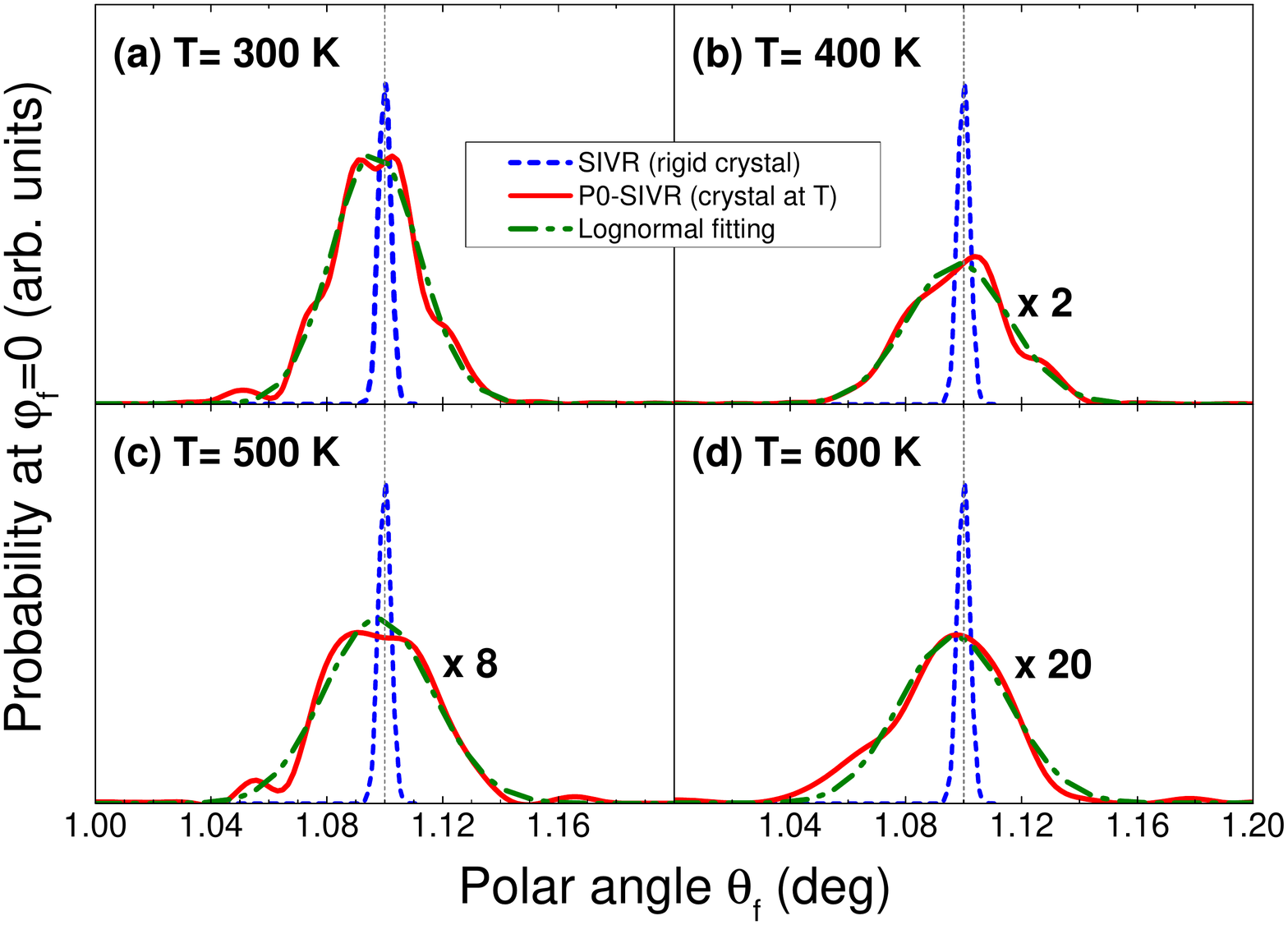} \centering
\caption{(Color online) Intensity profile of the central maximum at $\protect%
\varphi _{f}=0$, as a function of the polar angle $\protect\theta _{f}$, for
the case of Fig. \protect\ref{map-0t300}, considering different
temperatures: (a) $T=300$ K, (b) $T=400$ K, (c) $T=500$ K, and (d) $T=600$
K. In all the panels, red solid line, differential probability derived
within the P0-SIVR approach; blue dashed line, SIVR probability for a rigid
crystal; green dot-dashed line, fitting of P0-SIVR results by means of a
lognormal distribution, as given by Eq.(\protect\ref{lognormal}). Vertical
gray dashed line, ideal $\protect\theta _{f}$- position on the Laue circle
(i.e., $\protect\theta _{f}=\protect\theta _{i}$).}
\label{profile-fi0}
\end{figure}

An important feature introduced by the thermal vibrations\ is the $\theta
_{f}$- dispersion of the GIFAD patterns, which transforms the punctual spots
produced by the rigid crystal into vertical streaks, as observed in Fig. \ref%
{map-0t300}. \ With the aim of analyzing the dependence on $T$ of such a
polar-angle spread, in Fig. \ref{profile-fi0} we display the polar profile
of the central maximum, that is, the differential probability $%
dP^{(P0-SIVR)}/d\theta _{f}$ at $\varphi _{f}=0$, as a function of the polar
angle $\theta _{f}$, for the same temperatures as in Fig. \ref{spectra-T}.
In each panel, P0-SIVR\ results including phonon contributions are compared
with the SIVR profile corresponding to a rigid crystal. In contrast with the
SIVR spectrum, which presents a sharp peak centered at the specular
reflection angle (i.e., $\theta _{f}=\theta _{i}$), the polar distribution
derived within the P0-SIVR approach shows a broad maximum, whose intensity
decreases as $T$ increases. This latter effect is only partially due to the
screening of the projectile-surface interaction introduced by the
Debye-Waller factor in Eq. (\ref{Vcrys}). Additionally, in Fig. \ref%
{profile-fi0} we observe that the width of the P0-SIVR\ peak is affected by
the thermal fluctuations, showing a slight increase as the temperature
augments.

Similar thermal spread of the polar angle $\theta _{f}$ was predicted by
Manson \textit{et al.} in Ref. \cite{Manson2008}, where the polar profile of
the GIFAD patterns was estimated as following a lognormal distribution. In
order to verify \ this behavior, we fit the $\theta _{f}$- profiles at $%
\varphi _{f}=0$ derived with the P0-SIVR approach for different $T$ values
with the lognormal function
\begin{equation}
\mathcal{P}(\theta _{f})=\frac{A}{\omega \ \theta _{f}}\exp [\frac{%
-2(ln(\theta _{f}/\theta _{c}))^{2}}{\omega ^{2}}],  \label{lognormal}
\end{equation}%
where $A$, $\theta _{c}$, and $\omega $ are fitting parameters that depend
on $T$. The resulting $\mathcal{P}(\theta _{f})$ functions, displayed with
green dot-dashed lines in Fig. \ref{profile-fi0}, reproduce the P0-SIVR
curves quite well, allowing as to determine the width $\omega $ of the
effective Gaussian distribution of Eq. (\ref{lognormal}) as a function of $T$%
.
\begin{figure}[tbp]
\includegraphics[width=0.5\textwidth]{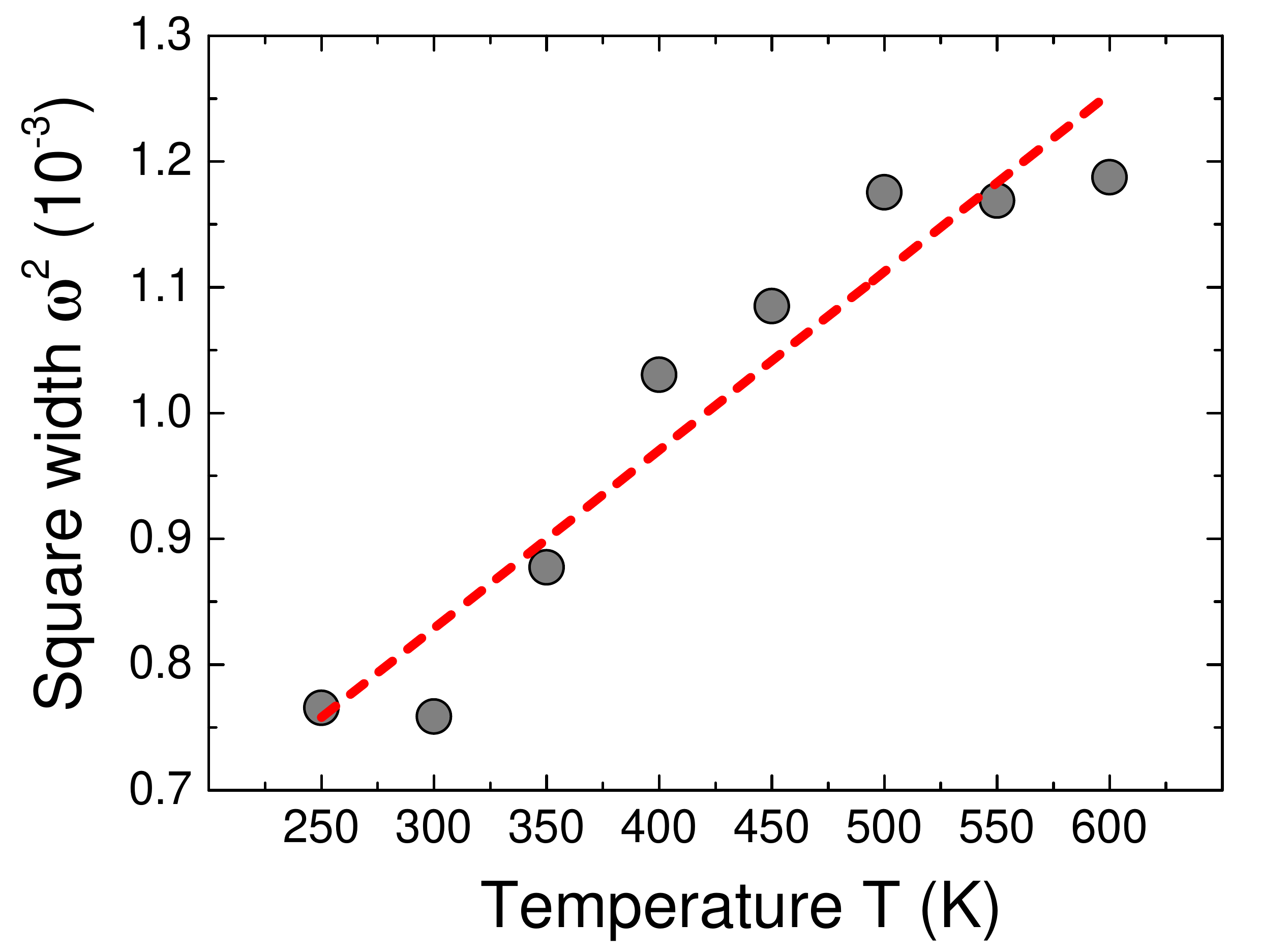} \centering
\caption{(Color online) Square width $\protect\omega ^{2}$ of the Gaussian
distribution given by Eq. (\protect\ref{lognormal}), as a function of the
temperature. Circles, values derived by fitting P0-SIVR results; red dashed
line, linear fitting of the present $\protect\omega ^{2}$ values. }
\label{ancho-T}
\end{figure}

Within the lognormal model of Ref. \cite{Manson2008}, the square width of
the Gaussian distribution given by $\mathcal{P}(\theta _{f})$ is obtained as
proportional to the mean-square vibrational amplitude normal to the surface
plane. That is, $\omega ^{2}$ $=\Gamma ^{2}\left\langle
u_{z}{}^{2}\right\rangle _{T}$, where the coefficient of proportionality $%
\Gamma $ coincides with the normalized slope of the projectile-surface
potential, which is assumed as $V_{0}\exp (-\Gamma Z)$ in Ref. \cite%
{Manson2008}, with $Z$ being the distance to the surface. Then, to test this
relation in Fig. \ref{ancho-T} we plot $\omega ^{2}$ values obtained by
means of the lognormal fitting of Eq. (\ref{lognormal}) for different
temperatures $T$ in the $250$-$600$ K range. Even though the points of Fig. %
\ref{ancho-T} show an appreciable dispersion, they seem to follow a linear
tendency, leading to a rate $\Gamma \approx 0.22$ \AA $^{-1}$. But this $%
\Gamma $ value is one order of magnitude smaller than the normalized slope
of the projectile-surface potential around the turning point, $\ -\left[
V_{PS}(Z)\right] ^{-1}dV_{PS}/dZ$, in contrast with the prediction by Manson%
\'{}%
s model \cite{Manson2008}.

\subsection{Experimental comparison}

\begin{figure}[tbp]
\includegraphics[width=0.4\textwidth]{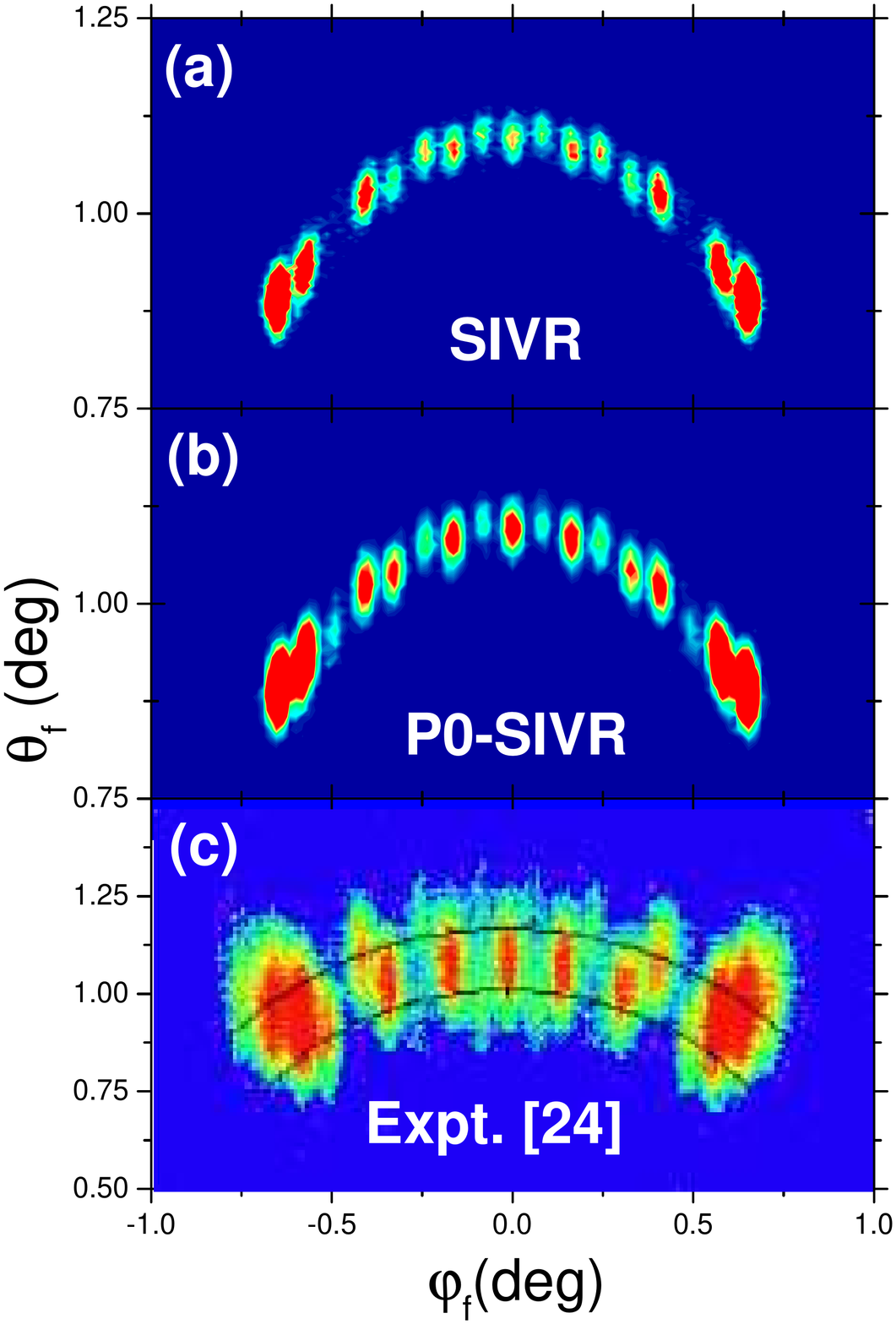} \centering
\caption{(Color online) 2D projectile distribution, as a function of $%
\protect\theta _{f}$ and $\protect\varphi _{f}$, for the case of Fig.
\protect\ref{map-0t300}. Results derived within (a) the SIVR approach for a
rigid crystal and (b) the P0-SIVR approximation, including thermal lattice
vibrations, both with the collimating parameters given in Sec. III.D; and
(c) experimental data extracted from Ref. \protect\cite{Schuller2009c}. }
\label{map-expto}
\end{figure}

Finally, we check the validity of present results for the He-LiF(001) system
by contrasting double differential P0-SIVR probabilities with experimental
data extracted from Ref. \cite{Schuller2009c}. In Fig. \ref{map-expto}, 2D
angular distributions as a function of $\theta _{f}$ and $\varphi _{f}$,
derived from the P0-SIVR and SIVR approaches, are compared with the
experimental intensity\ distribution as recorded with\ a position sensitive
detector \cite{Schuller2009c}. In this case, P0-SIVR and SIVR simulations
were done by assuming an atomic beam collimated through a square slit of
size $d=0.3$ mm, placed at a distance $L=25$ cm from the surface. Although
details about the collimation setup were not provided in Ref. \cite%
{Schuller2009c}, the chosen collimating parameters agree with those reported
in other articles by the same group \cite{Seifert2015}.

From Fig. \ref{map-expto}, the\ P0-SIVR distribution shows a good accord
with the experimental data, both presenting a central region with even Bragg
orders much higher than the odd ones. Instead, the relative intensities of
the Bragg peaks provided by the SIVR approximation differ from the formers,
showing a less pronounced intensity contrast between even and odd Bragg
orders around the axial direction. In addition, in a similar fashion to Fig. %
\ref{map-0t300}, the effect of thermal fluctuations gives rise to an
increase of the polar spread of the P0-SIVR patterns, with respect to that
of the SIVR distribution. However, note that the P0-SIVR distribution [Fig. %
\ref{map-expto} (b)] \ has a smaller $\theta _{f}$- dispersion than the
experiment [Fig. \ref{map-expto} (c)]. It suggests that the experimental
polar spread might have additional contributions due to inelastic processes
involving phonon transitions or crystal defects, both mechanisms not
included in the present model.

\section{Conclusions}

In this work the influence of temperature on GIFAD patterns for the
He-LiF(001) system has been studied by using the P0-SIVR approximation. The
P0-SIVR approach is a semiquantum method that describes zero-phonon
scattering including the contribution of thermal lattice vibrations. These
thermal vibrations were found responsible for the polar spread of the
diffraction patterns, which transforms the sharp Bragg maxima produced by
the rigid crystal into vertical streaks. Furthermore, P0-SIVR spectra as a
function of the azimuthal angle vary with the temperature, which strongly
modifies the relative intensity of the diffraction maxima, while the
azimuthal width of the peaks is slightly affected by the thermal lattice
fluctuations. As it happens for semiconductor surfaces, well defined GIFAD
patterns are obtained for temperatures as high as $600$ K. In addition, by
analyzing the polar profile of the central Bragg maximum as a function of $T$%
, the lognormal behavior proposed in Ref. \cite{Manson2008} was also
scrutinized. We found that the square width $\omega ^{2}$ of the \textit{%
effective} lognormal distribution roughly increases linearly with the
crystal temperature, but the slope is much lower than that estimated in Ref.
\cite{Manson2008} from a simple potential model.

Present P0-SIVR results were contrasted with the experimental projectile
distribution from Ref. \cite{Schuller2009c}, showing an overall good
agreement. Nevertheless, the polar extension of the P0-SIVR pattern
underestimates that of the experiment, suggesting that other effects, like
phonon excitations or surface defects, might contribute to the polar
dispersion of GIFAD patterns at room temperature. Therefore, an exhaustive
experimental study of the $T$ dependence of GIFAD from insulator surfaces\
should be desirable.

\begin{acknowledgments}
The authors acknowledge financial support from CONICET and ANPCyT of
Argentina.
\end{acknowledgments}

\bibliographystyle{unsrt}
\bibliography{HeLiF-temper}

\end{document}